# Spherically confined isotropic harmonic oscillator

K.D.Sen*
School of Chemistry
University of Hyderabad
Hyderabad-500 046 , India

and

Amlan K. Roy
Department of Chemistry
University of California Los Angeles (UCLA)
Los Angeles 90095-1569

**Abstract**

The generalized pseudospectral Legendre method is used to carry out accurate calculations of eigenvalues of the spherically confined isotropic harmonic oscillator with impenetrable boundaries. The energy of the confined state is found to be equal to that of the unconfined state when the radius of confinement is suitably chosen as the location of the radial nodes in the unconfined state. This incidental degeneracy condition is numerically shown to be valid in general. Further, the full set of pairs of confined states defined by the quantum numbers [(n+1, $\ell$); (n, $\ell$+2)], n = 1,2,.. , and with the radius of confinement $\{(2\ell+3)/2\}^{1/2}$ a.u. ,which represents the single node in the unconfined (1, $\ell$) state, is found to display a constant energy level separation exactly given by twice the oscillator frequency. The results of similar numerical studies on the confined Davidson oscillator with impenetrable boundary as well as the confined isotropic harmonic oscillator with finite potential barrier are also reported .The significance of the numerical results are discussed .

*sensc@uohyd.ernet.in

**I. Introduction**

The idea of a spherically confined atom was invented by Michels et al [1] by enclosing the hydrogen atom inside an impenetrable spherical cavity. During the recent years, successful laboratory realization of a variety of new experimental systems of specific interest as the future nano and molecular electronic circuit devices, including quantum computers , has generated tremendous interest in the development of the idea of quantum mechanical confinement . The experimental systems e.g. , two and three dimensional graphite cones and carbon nanotube rings [2-3] , InGaAs torus shaped nano-rings [4-6] and carbon nanotube rings [7-8] present themselves as challenging problems in modern quantum chemistry .In the literature dealing with the theoretical studies involving confinement models , both hard and soft confining boundaries with a variety of shapes and sizes have been considered [9] .The majority of theoretical studies have been centered around the effect of confinement on energy levels of the unconfined system. In particular, it has been observed that the degeneracy and relative ordering of energy levels of the unconfined system are both influenced significantly under the effect of confining potentials. Considerable theoretical efforts have recently been made in performing more accurate computations on simple model systems like the hydrogen[10] and helium atom[11] which could also serve as a benchmark for the approximate methods. For all angular momentum values, the exactly solvable problems of the free (unconfined) hydrogen atom (FHA) and the free isotropic harmonic oscillator (FIHO), respectively, define the standard text book problems in quantum mechanics .Under the conditions of spherical confinement, on the other hand, while the spherically confined hydrogen atom (SCHA) has been studied in considerable detail, the spherically confined isotropic harmonic oscillator (SCIHO) remains to be studied similarly. The so called accidental degeneracy of the free hydrogen atom (FHA) and the free isotropic harmonic oscillator (FIHO), respectively, is generally understood [12-17] in terms of the corresponding symmetry groups SO(4) and SU(3) , respectively. The accidental degeneracy of these systems is characterized by different sets of parity (even or odd $\ell$ ) conditions. It is interesting to note that the ground state sum of the Shannon information entropy [18-19] of the density in the position and momentum space is larger for the FHA than the FIHO with the latter representing the minimum Heisenberg uncertainty. It is well known that in both cases the accidental degeneracy is broken under the spherical confinement conditions. For the SCHA, two kinds of additional degeneracy can be defined[20-21] both of which result from the specific choices of the radius of confinement $R_c$ chosen exactly at the radial nodes corresponding to the FHA wave functions . Firstly ,for all $n \geq \ell + 2$ , the SCHA state denoted by (n, $\ell$) is degenerate with (n+1, $\ell$+2) state when both of then are confined at $R_c$ [= ($\ell$+1) ($\ell$+2) ] which defines the radial node in the free ($\ell$+1, $\ell$) state .Accordingly, the confined state energies [$E_{2s}(R_c=2)$, $E_{3d}(R_c=2)$] , [$E_{3s}(R_c=2)$, $E_{4d}(R_c=2)$] ,...; [$E_{3p}(R_c=6)$, $E_{4f}(R_c=6)$] , [$E_{4p}(R_c=6)$, $E_{5f}(R_c=6)$], ... are degenerate. This kind of degeneracy for the SCHA is referred to as the *simultaneous degeneracy*. Secondly, corresponding to $R_c$ = ($\ell$+1) ($\ell$+2) , the confined ($\ell$+1, $\ell$) state is found to be iso-energic with ($\ell$+2) state of the FHA with energy – 1/{2($\ell$+2)$^2$ } a.u. This degeneracy is referred to as the *incidental degeneracy* .The simultaneous and incidental degeneracies have been shown to result from the Gauss relationship applied at a unique radius of confinement by the confluent hypergeometric functions which describe the general solutions $\psi(\mathbf{r})$ of confined hydrogen atom problem . For a detailed numerical study on this subject we refer to the recent work [22].

To our knowledge, the numerical study of the incidental degeneracy of the ground state of the SCIHO with $R_c$ chosen as the first excited (1,0) state of zero angular momentum corresponding to the FIHO *alone* has been discussed so far in the literature [23-33]. In a recent proposal on quantum circuits, a universal and deterministic utilization of the *confined* harmonic oscillator states have been proposed [34]. In the light of such widespread applications it seems highly desirable to examine the level ordering of the SCIHO along the lines considered recently [22] for the SCHA.

The purpose of this paper is twofold : (a) to test ,with appropriate modifications, the ability of the generalized pseudospectral (GPS) Legendre method with mapping procedure pioneered by Chu and coworkers [35-37] to describe the SCIHO and SCHA systems , and (b) to numerically study the characteristic features in the ordering of energy levels of the SCIHO when the radius of confinement, $R_c$ , is specifically chosen as that corresponding to the location of radial nodes in the FIHO .

The lay out of this paper is as follows .In section II, we give a brief outline of the GPS method used in this work and show specifically how the method can be conveniently adapted to handle the spherically confined systems. In section III, we present the results of our numerical calculations on the test cases of the SCIHO and SCHA and compare them with the most accurate results available in the literature. In the same section, the results of more extensive studies on the SCIHO are presented and discussed .Finally, a summary of the main conclusions is given in section IV.

**II. Methodology**

One of us has recently applied the GPS method to study a variety of singular potentials [38] in the unconfined state. In this section, we present only a brief outline of the GPS method employed to solve the radial eigenvalue problem with central potentials. A more detailed account can be found in the available literature [35-38].

Without loss of generality, the desired radial Schrödinger equation can be written as (in atomic units unless otherwise states)

$$\hat{H}(r)\phi(r) = \varepsilon\phi(r) \ , \qquad (1)$$

where the Hamiltonian includes the usual kinetic and potential energy operators,

$$\hat{H}(r) = -\frac{1}{2}\frac{d^2}{dr^2} + V(r) \ , \qquad (2)$$

with

$$V(r) = \frac{1}{2}r^2 + \frac{\ell(\ell+1)}{2r^2} + V_c(r) \qquad (3)$$

where $V(r)$ in Eq. (3) includes the standard harmonic oscillator potential with the force constant k = 1 and an additional confining potential, $V_c(r)$. Our model of spherical confinement is shown in Fig. 1 which also defines the confining potential $V_c(r)$ given in Eq. (3). The calculations for the SCIHO confined inside an impenetrable boundary is carried out [37] by setting the maximum radial distance, $r_{max}$, as $R_c$, the radius of spherical confinement. With reference to Fig. 1, this corresponds to $R_a = 0$ and $R_b = R_c$. The wave function at $R_c$ thus satisfies the Dirichlet boundary condition $\psi(0) = \psi(R_c) = 0$ and the confined wave function $\psi(r)$ is defined over $0 \leq r \leq R_c = R_a$. The usual finite-difference spatial discretization schemes often require a large number of grid points to achieve good accuracy, because the majority of these methods employ a uniform mesh[22]. The GPS method, however, can give *nonuniform* and optimal spatial discretization accurately. This allows one to work with a denser mesh at shorter *r* regions and a coarser mesh at large *r*. Additionally the GPS method is computationally orders of magnitude faster than the finite-difference schemes. One of the principal features of the GPS scheme is that a function *f(x)* defined in the interval $x \in [-1, 1]$ can be approximated by the polynomial $f_N(x)$ of order *N* so that

$$f(x) \cong f_N(x) = \sum_{j=0}^{N} f(x_j) g_j(x) , \qquad (4)$$

And the approximation is *exact* at the *collocation point* $x_{j'}$ i.e.,

$$f_N(x_j) = f(x_j) . \qquad (5)$$

In this work, we employ the Legendre pseudospectral method using $x_o=-1$, $x_N=1$, where $x_j$ *(j=1,...,N-1)* are obtainable from the roots of the first derivative of the Legendre polynomial $P_N(x)$ with respect to *x*, i.e.,

$$P_N'(x_j) = 0 . \qquad (6)$$

The cardinal functions *g(x)* in Eq.(6) are given by the following expression:

$$g_j(x) = -\frac{1}{N(N+1)P_N(x_j)} \frac{(1-x^2)P_N'(x)}{x - x_j} , \qquad (7)$$

Obeying the unique property $g_j(x_{j'})=\delta_{j'j}$. Now the semi-infinite domain $r \in [0, x]$ is mapped into the finite domain $x \in [-1,1]$ by the transformation *r=r(x)*. One can make use of the following algebraic nonlinear mapping:

$$r = r(x) = L\frac{1+x}{1-x+\alpha} , \qquad (8)$$

where $L$ and $\alpha=2L/r_{max}$ may be termed as the mapping parameters. Finally, the transformed Hamiltonian is solved by introducing the relation,

$$\psi(r(x)) = \sqrt{r'(x)}\ f(x) \qquad (9)$$

along with the symmetrization procedure.

The performance of the method has been tested earlier[35-38] with respect to the variation of the mapping parameters with excellent numerical performance for a variety of unconfined ($R_c = \infty$) potentials. The FIHO eigenvalues show the well known "$2n + \ell$" degeneracy of the given $N^{th}$ (=$2n + \ell$) energy level with the equidistant eigenvalues defined by $(2n + \ell + 3/2)\ \hbar\omega$. Accordingly, the observed level ordering is given in $(n, \ell)$ notations, by : (0,0), (0,1), [ (1,0), (0,2) ], [(1,1), (0,3) ], … with energy as 1.5, 2.5, 3.5, 4.5, … oscillator units ($\hbar\omega$) of energy, respectively. We shall denote these levels as 1s, 1p, [ 2s, 1d ], [ 2p, 1f ], [ 2d, 1g] … etc. We note here that the level spacing between the successive states having the same $\ell$ values is given by $2\hbar\omega$ which includes the $\Delta\ell = 2$ pairs [ 2s, 1d ], [ 3s, 2d ], …; [ 2p, 1f ], [ 3p, 2f ],…etc. We shall retun to this characteristic feature of the FIHO spectrum in section III (b). For all the calculations done in this work, a consistent set for the numerical parameters for the *unconfined* state ($r_{max}=200$, $\alpha=25$, and $N=300$) has been used, which seem to be appropriate for the current problem. In addition to confinement condition $\psi(0) = \psi(R_c)=0$ introduced above we have also considered the confinement within the concentric spheres [22] by identifying the semi-infinite domain $r \in [0, x]$ with $r \in [R_a, R_b]$ to correspond to the requirement that $\psi(R_a) = \psi(R_b)=0$. In this case, r is bounded according to $R_a \leq r \leq R_b$ and the wave function is restricted inside the spherical shell as shown in Fig. 1. We further note here that GPS method is equally suitable for the Neumann boundary condition where a finite (soft) potential wall is employed and we shall report one set of such calculations on the SCIHO towards the end of section III.

**III. Results and Conclusions**

We first report the results of the GPS calculations of energy levels of the SCIHO and the SCHA inside an impenetrable wall and compare them with the most accurate previous estimates .In Table 1 we have listed the results of our calculations on the SCIHO and compared them with very accurate estimates [29] reported recently. The two sets of results are found to be in excellent agreement .In Table 2 , we have similarly compared the results of our calculations of the energy levels of the SCHA with the most accurate numerical data reported earlier [10] . Once again , the two sets of results agree almost exactly with each other .It is concluded from Table 1-2 that the GPS method is able to describe the energy levels of the spherically confined systems with remarkable accuracy, as found earlier also for the unconfined systems[35-38] .The presently calculated energies are significantly more accurate than all the other hitherto reported results for a large number of n , $\ell$ , $R_c$ values .Consequently, the scope of application of the GPS method is hereby extended to cover a large variety of the problems in chemical physics enumerated by Chu [35] from free to the confined systems with equal ease and accuracy . In the subsequent Tables 4-7 , we shall report the results only up to five digits following the decimal as they are appropriate enough to support the conclusions drawn .

**(a) Incidental degeneracy**

We present our results on the *incidental degeneracy* of the SCIHO. In Table 3, we have listed the radial nodes for up to n=3, $\ell=3$, wave functions of the free isotropic harmonic oscillator, all derived numerically from the reported spherical Hermite wave functions [39]. The radial node in the first excited $(1,\ell)$ state of the FIHO is given by $\{(2\ell+3)/2\}^{1/2}$ a. u. with a scale factor given by $1/(k^{1/4})$ in a. u. We shall report the calculations based on the choice of k = 1. Our calculations have been carried out using the $R_c$, the radius of impenetrable sphere as one of the nodal points given in Table 3. It is well known that the so called accidental degeneracy is broken under the spherical confinement with the higher $\ell$ values becoming lower in energy for a given n value. Under spherical confinement, the energy of the (0,0) state increases with decreasing $R_c$. With $R_c$ coinciding with the node corresponding to the unconfined (1,0) state with energy $3.5\,\hbar\omega$, the energy of the confined (0,0) state also attains the same value. This is due to the fact that, apart from normalization, the unconfined (1,0) state truncated at the node, is a solution of the (0,0) state confined at the specified node given by $R_c = (3/2)^{1/2}$. In Table 4, we show numerically how the energy of the SCIHO state given by $(n, \ell = 0)$ can be made to coincide with that corresponding to any other higher $(n^*, \ell = 0)$ *unconfined* state with $n^* > n$ energy by suitably choosing $R_c$ at the radial nodes of the unconfined state. For example, we define three SCIHO states by choosing $R_c$ as 0.816288 ($R_{c1}$), 1.673552 ($R_{c2}$) and 2.651961 ($R_{c3}$) a.u., which define three consecutive nodes of the unconfined (3,0) state with n*=3. They are denoted as the (0,0) (1,0), and (2,0) states confined at $R_{c1}$, $R_{c2}$, and $R_{c3}$, respectively. All of these confined states correspond to the energy equal to $7.5\,\hbar\omega$ which coincides with the energy of the state (3,0) of the FIHO. In Fig. 2, this particular example has been displayed schematically on the left hand side with the confined states denoted by $1s_c$, $2s_c$ and $3s_c$, respectively. Similarly, on the right hand side in Fig. 2, another set of three confined states, all with the common final radius given by $R_b = \infty$, but different $R_a$ values defined by the three successive radial nodes are shown. The set of six SCIHO states in Fig.2 individually correspond to the same energy 7.5 $\hbar\omega$ i.e. the energy of the 4s state of the FIHO. We have also verified that the similar incidental degeneracy generally holds good for $\ell = 1$-3. As a representative example, in Table 5, we have listed the SCIHO states corresponding to $\ell = 3$ showing the occurrence of incidental degeneracy. For the excited states in particular where the approximate variational and/or perturbation methods are employed, the incidental degeneracy condition can be used as a criterion for evaluating the quality of approximate wave functions. It is significant to note that the locations of the nodes in the unconfined state scale as $1/(k^{1/4})$, k being the force constant. Consequently, the scale of the radius of confinement showing incidental degeneracy could be made to vary from the nuclear radii to the sizes of large clusters of atoms and molecules and the GPS method can be employed at the corresponding force constant.

**(b) Confined isotropic harmonic oscillator states with $\Delta\ell = 2$**

In this section, the behavior of the SCIHO under the confinement conditions specified for the simultaneous degeneracy of the SCHA system will be discussed. In particular, the pairs of the SCIHO states with $\Delta\ell = 2$ which are both confined at the same $R_c = \{(2\ell+3)/2\}^{1/2}$ a. u. will be studied. We begin by noting that the simultaneous degeneracy condition for the SCIHO states is already met under the *unconfined* condition represented by the FIHO defined by the boundary conditions $[\psi(R_i) = \psi(R_c) = 0; R_i = 0, R_c = \infty]$. It is therefore of interest to examine if at any other finite radius of confinement there exists a level ordering for the pair states defined by $\Delta\ell = 2$ of the SCIHO which is , in particular ,characterized by the oscillator energy unit(s) of $\hbar\omega$ . In Table 6 , we have collected the results of our calculations corresponding to the *two* different sets comprising of the pairs of SCIHO states defined by $\Delta\ell = 2$ which are *degenerate* under the unconfined condition at the $N^{th}$ level : (a) $[(1,0) , (0,2) ]_{N=2}$ , $[(2,0) , (1,2) ]_{N=4}$  , and (b)  $[(1,5) , (0,7) ]_{N=7}$ , $[(2,5) , (1,7) ]_{N=9}$ , where   $R_c$ has been chosen as  $\{(2\ell+3)/2\}^{1/2}$  with $\ell=0 , 5$  , respectively. While only the representative two sets corresponding to the (s, d) and (h, j) pairs are compiled in Table 6 , our calculations show that *all* sets of such $\Delta\ell = 2$ pairs : [ (2s , 1d) , (3s,2d) , … ] ; [ (2p,1f) , (3p,2f) , .. ] ; [ (2d, 1g) , (3d,2g)…] ; **…** at their *respective* given radius of confinement display a *uniform* energy separation of $2\hbar\omega$ .

In Fig. 3 , we have displayed $\Delta E = E(n,d) - E(n+1,s)$ corresponding to the pairs (2s,1d) and (10s,9d) , respectively , for the SCIHO at different $R_c$ values. At $R_c = 1.2247448$ a.u. , i.e. the only node in the 2s state of the FIHO , the two pairs correspond to $\Delta E = 2\hbar\omega$. This makes an interesting comparison of the SCIHO with the corresponding SCHA ($-1/r$ vs $r^2$ , *confined* potential with impenetrable walls). As discussed earlier one obtains[20-22] the *simultaneous degeneracy* condition for the SCHA where the pair of states (2s,3d) , (3s,4d) **...** in the unconfined condition (FHA) are *originally separated* according to energy defined by $-1/2n^2$ a. u. Interestingly, while the pair of states of the FIHO given by $[(n+1 , \ell ) ; (n, \ell +2)]$ ,   n = 1,2,**…** ,are *originally degenerate*, the confined states at the chosen radius of confinement produce an entire set of oscillator state pairs uniformly separated by $2\hbar\omega$ .As noted at the end of section II above, the quantity $2\hbar\omega$ is the hallmark of the FIHO spectrum which defines the equidistant $\Delta E$ corresponding to the successive levels of a given $\ell$ value .It has been conjectured that the simultaneous degeneracy in the SCHA is due to some hidden symmetry. The numerical results in Table 6 indicate the presence of a different kind of hidden symmetry in the SCIHO problem which generates the simultaneous degeneracy at $R_c = \infty$ while displaying the $\Delta E = 2\hbar\omega$ condition stated above at $R_c = \{(2\ell+3)/2\}^{1/2}$ a. u. i.e. the radial node of the first excited $(1,\ell)$ of the unconfined isotropic harmonic oscillator. This calls for an independent special analysis based on the group algebraic methods. We refer to the special confinement effect observed at $R_c = \{(2\ell+3)/2\}^{1/2}$ as the *simultaneous frequency doubling* effect. In Fig. 4, we have plotted the confined 10s and 9d density at $R_c = 1.224788$ a.u. The compact distribution in case of 9d state represents the energy lowering relative to the 10s state. In a large number of cases pertaining to the confinement studies involving the anharmonic potentials, the harmonic oscillator states are employed as the zeroth order approximation in a perturbation expansion. The numerical results reported here have important bearings on such studies. A similar situation also holds good for the confined singular potentials including the multiple well potentials. In addition, as in the case of *incidental* degeneracy, the uniform energy separation of $2\hbar\omega$ for the confined states defined by $\Delta\ell = 2$ can be used as benchmark for the evaluation of the approximate quantum mechanical models dealing with confined oscillator potentials in general .

In the light of above results, it is interesting to examine the Davidson oscillator defined by the potential [40]

$$V(r) = \frac{1}{2}r^2 + \frac{\ell(\ell+1)}{2r^2} + \frac{\lambda}{2r^2} \qquad [10]$$

which has been found useful in describing the rotational-vibrational states of diatomic molecules[41-42]. Further, a five dimensional version of Davidson oscillator has been considered as a model for nuclear rotations and vibrations[42] .For the three dimensional isotropic Davidson oscillator , the energy levels are exactly given by

$$E_{n\ell} = \left[2n+1+\{(\ell+\frac{1}{2})^2+\lambda\}^{1/2}\right]\hbar\omega \qquad [11]$$

The substitution of $\ell$ by its enhanced effective value $\ell_{eff} = \frac{1}{2}+\{(\ell+\frac{1}{2})^2+\lambda\}^{1/2}$ in $\{(2\ell+3)/2\}^{1/2}$ represents the corresponding location of the nodes of the unconfined Davidson oscillator states. The (2s,1d) degeneracy observed for the free isotropic harmonic oscillator is already lifted in the case of the *unconfined* Davidson oscillator with the 1d level becoming relatively more stable. Indeed , a generalization of the isotropic harmonic oscillator potential given by $Br^2 + A/r^2$ , with A and B as constants , expressed as the Gol'dman and Krivchenkov Hamiltonian has been recently used to represent the unperturbed part of a class of anharmonic singular potentials[43] .The existence of exact solutions corresponding to the unconfined D-dimensional Gol'dman and Krivchenkov Hamiltonian suggests that the spherically confined calculations of energy derived from such potentials with special attention on the inter-dimensional degeneracy[44] would make an interesting study. Similarly, employing the $\ell_{eff}$ in each case it is possible to extend the present study on the degeneracy characteristics to the Davidson oscillator. It is interesting to note that both the terms A and B (through $\ell_{eff}$ ) now determine the locations of nodes in the free Davidson oscillator. In Table 7, we have presented the result of our numerical test of the incidental degeneracy for the $\ell$ =0 ground state of the Davidson oscillator with $\lambda$ =1. The first 10 levels of the unconfined oscillator are displayed under column 1. For $\ell$ =0 , the $R_c$ is calculated as $\{(2\ell_{eff}+3)/2\}^{1/2}$ which is noted under column 3 .Owing to the extra repulsive term in the Davidson potential , the occurrence of the node at 1.45535 a.u. is found to be larger than that of the free isotropic harmonic oscillator at 1.22474 a.u. .As shown in Table 7, the incidental degeneracy is confirmed by finding the energy of the confined state (entered as the first entry under column 4) becoming equal to the unconfined Davidson oscillator of the first excited level (1,0) . There is another notable feature observed for the pair of states defined by $\Delta\ell = 2$ . As shown in last three columns, the difference of $\Delta E$ , between the successive pairs of $\Delta\ell = 2$ states is found to decrease rather slowly as the number of energy level increases. We have found this to be valid for $\Delta\ell = 2$ states of the Davidson oscillator, in general.

Finally, we conclude this report by demonstrating that the GPS method is also suitable for carrying out confined state calculations with *finite* potential barrier. The radial location of the wall can be conveniently chosen arbitrarily as one of the points on the variable $r_i$ grid, and a fixed potential $V_c(R_c)$ defined over the entire range $r_i = R_c$ to $r_{max}$ can be incorporated in the calculations. In Table 8, we present the variation of the 1s ground state of the SCIHO corresponding to three different choices of $R_c$ given by 1.22511 , 1.99975 and 4.00052 a.u. , respectively. The first choice is closest in the chosen radial grid to the location of the first node in (1,0) state of the FIHO. The other two $R_c$ values are chosen as nearest to 2 a.u. and 4 a.u. respectively .Four different penetrable (soft) potential barriers, $V_c(R_c)$ , given by 100, 50, 10 and 5 a.u , respectively and located at each of the three $R_c$ above have been considered .In Table 8 , the corresponding confined energy values at a given $R_c$ are denoted by E' ($R_c$=1.22511), E" ($R_c$=1.99975), and E''' ($R_c$=4.00052), respectively. These results suggest that the energy levels of the SCIHO are very sensitive to the magnitude of the confining potential $V_c(R_c)$ . We are presently carrying out a more detailed calculations on this system including other polynomial model potentials as well as the ground and excited states of some light atoms using the GPS method.

## 4. Summary

In summary, using the generalized pseudospectral Legendre method , high precision numerical calculations of eigenvalues of the SCHA and the SCIHO with impenetrable boundaries (Dirichlet boundary condition) have been reported. When the radius of confinement is *specifically* chosen as the location of the radial nodes in the unconfined state, the energy of a set of the suitably confined SCIHO states can be made to coincide with that of the unconfined state. This *incidental* degeneracy condition is shown to be fulfilled in general. As the energy values corresponding to the FIHO potential are *exactly* known, the incidental degeneracy condition can be used to test the accuracy of the approximate methods prescribed for the SCIHO system.  The pairs of confined states defined by $[(n+1, \ell) ; (n, \ell+2)]$ , n = 1,2, … and with the radius of confinement $R_c = \{(2\ell+3)/2\}^{1/2}$ a.u. are found to display , for all n , a constant separation of energy *exactly* given by $2\hbar\omega$ i.e. twice the oscillator frequency. This is in interesting contrast with the similarly confined hydrogen atom problem[20-22] where the *unconfined* ($R_c = \infty$) pairs of states $[(n, \ell) ; (n+1, \ell+2)]$ , n = 2,3… with different energy values , assume degeneracy , for all n , at $R_c = (\ell+1)(\ell+2)$ a.u. .The simultaneous frequency doubling effect obtained under the spherical confinement is the central result obtained in this work and it has implications in several related oscillator potentials[45-46]. This condition can also be used as benchmark for the approximate analytic methods proposed to study the SCIHO states. The specific case of Davidson oscillator has been similarly discussed . The GPS method has been extended to the confined isotropic harmonic oscillator satisfying the Neumann boundary condition with finite potential barrier. It appears that both the SCHA and SCIHO models display interesting energy degeneracy effects which arise from the specific choice of the radius of confinement as the locations of the radial nodes in the FHA and FIHO , respectively.

## Acknowledgement

KDS is grateful to Dr. H. E. Montgomery Jr and Dr. N. A. Aquino for the constant motivation and encouragement throughout the course of  this work .He  thanks the Department of Science and Technology , New Delhi , for financial support.

**References** :

[1] A.Michels, J. de Boer , A. Bijl,  Physica   **4** , 981 (1937 ) .

[2] G. Cuniberti , J. Yi, and M. Porto, Appl. Phys. Lett. **81**,  850 (2002).

[3] G. Zhang , X. Jiang,  and E. Wang , Science **300**, 472 (2003).

[4] J. M. García, G. Medeiros-Ribeiro, K. Schmidt, T. Ngo, J. L. Feng, A. Lorke, J. Kotthaus, and P. M. Petroff , Appl. Phys. Lett. **71**, 2014 (1997).

[5] A Lorke, R. J. Luyken, A. O. Govorov,  J. P. Kotthaus , J. M. Garcia ,and P. M. Petroff Phys. Rev. Lett. **84**, 2223 (2000).

[6] A. Emperador ,M.  Pi , M. Barranco, and A. Lorke,  Phys. Rev. B **62**, 4573 (2000).

[7] J. Liu, H. Dai, J. H. Hafner, D. T. Colbert, R. E. Smalley, S. J. Tans and C. Dekker Nature (London) **385**, 780 (1997).

[8] R. Martel ,  H. R. Shea and  P. Avouris , Nature **398**, 299 (1999).

[9]  F.M.  Fernández ,  E.A.  Castro ,KINAM**4,**193(1982); P.O. Fröman ,S. Yngve, N. J. Fröman, J.  Math. Phys. **28 ,** 1813 (1987)1813 ; S. J. Yngve, J. Math. Phys.  **29**,931 (1988) ; W. Jaskólski,  Phys. Rep.  **271,** 1 (1996) ; A.L. Buchachenko , J. Phys. Chem **. 105 ,**5839 (2001) ; V.K. Dolmatov ,A.S. Baltenkov, J.P. Connerade , and S.T. Manson , Radiat. Phys. Chem.  **70**, 417 (2004). J.H.M. Lo, M. Klobukowski ,  G.Diercksen , Adv. Quantum. Chem. (2005) . (In Press) ;J. Gravesen , M. Willatzen and L.C. Lew Yan Voon , Phys. Scr. **72** ,105 (2005) .

[10] B.L.Burrows and M. Cohen ,Intl. J. Quantum Chem. **106**, 478 (2005) ; Phys. Rev. **A 72** ,032508 (2005) .

[11] N.A. Aquino, J.Garza , A. Flores-Riveros , J. F. Rivas-Silva , K.D. Sen , J. Chem. Phys. (2005) In Press.

[ 12 ] P.W. Higgs , J. Phys. **A 12** , 309 (1979 ) . References therein; R.W.Shea and P.K. Aravind , Am. J. Phys. **64** , 430 (1996) .For a lucid account of degeneracies of spherical well , harmonic oscillator and hydrogen atom in arbitrary dimensions.

[13] G.A.Baker ,Jr. ,Phys. Rev. A 103 ,1119 (1956 ) .

[14] D.M. Fradkin , Am. J. Phys. **33** , 207 (1965).

[15] V. Fock , Z. Phys. **98** , 145 (1935)

[16] S.P. Alliluev , Sov. Phys. JETP **6** ,156 (1958) .

[17] Z. Wu and J. Zeng , Phys. Rev. A 62 , 032509 (2000 ) .

[18] R.J. Yanez , W.Van Assche and J.S. Dehesa , Phys. Rev. **A 50** , 3065 (1994); J.S. Dehesa , A. Martinez-Finkelshtein and J. Sanchez-Ruiz , J. Comp. Appl. Math. **133** ,23 (2001).

[19] I.Bialynicki-Birula and J. Mycielski , Commun. Math Phys. **44** , 129 (1975) .

[20] A.V.Sherbinin , V.I. Pupyshev and A. Yu. Ermilov , Physics of Clusters, World Scientific , Singapore , 1997 , pp 217.

[21] V.I.Pupyshev and A.V. Sherbinin,Chem. Phys. Lett. **295** , 217 (1998) ; Phys. Lett. **A 299,**371 (2002) .

[22] K.D. Sen , J. Chem. Phys. **122** ,1943241 (2005)

[23] M. Moshinsky , The Harmonic Oscillator in Modern Physics: From Atoms to Quarks (London: Gorden and Breach , 1969).

[24]   D.S. Kothari and F. C. Auluck ,Science and Culture  **6** ,370 (1940) ; F. C. Auluck , Proc. Natl. Inst. Sci. India **7** , 133 (1941) ; ibid , **7** ,383 (1941) .

[25] S. Sengupta and S. Ghosh ,  Phys. Rev.  **115**  1681 (1959) .

[26] K K Singh , *Physica* **30** , 211 (1964) .

[27] F. M. Fernandez and E. A. Castro , Phys. Rev. **A  24** , 2883 (1981) ; G.A. Arteca, S.A. Maluendes , F. M. Fernandez and E. A. Castro , Intl. J.  Quantum Chem.  **24**,  169 (1983).

[28] J.L. Martin and S.A. Cruz , Am. J. Phys. **59** , 931 (1991) .

[29] N.A. Aquino , J. Phys. **A 30** ,2403 (1997) .

[30] R. Dutt , A. Mukherjee and Y. P. Varshni , Phys. Rev. **A 52** , 1750 (1995) .

[31] A. Sinha , J. Math . Chem. **34**, 201 (2003) .

[32] E. D. Filho and R. M. Ricotta , Phys. Lett **A 320** , 95 (2003) .

[33] P. Koscik and A Okopinska , J. Phys. **A 38** , 7743 (2005) .

[34] M.F. Santos , Phys. Rev. Lett. **95,** 010504  (2005) .

[35] S-I  Chu , J. Chem. Phys. **123** , 062207 (2005) .References therein.

[36] D. Telnov and S-I Chu , Phys. Rep. **390** , 1 (2004) .

[37] T. F. Jiang ,X-M. Tong and Shih-I Chu , Phys. Rev. **B 63** , 045317(2001) .

[38] A. K. Roy , Intl. J. Quant. Chem. **104** , 861 (2005) ; Phys. Lett. A  (2004), **321**,231 (2004); J. Phys. **G 30** ,269 (2004) .

 [ 39] D.B.Beard and O.B. Beard Quantum Mechanics with Applications , Allyn and Bacon Boston (1970) pp115

[40] P. M. Davidson , Proc. R. Soc . **135** , 459 (1932) .

[41 ] D.J. Rowe and C. Bahri , J. Phys. A **31** , 4947 (1998) .

[42 ] D. J. Rowe , J. Phys. **A 38** , 10181 ( 2005) .

[43] N . Saad , R. L. Hall , Q.D. Katatbeh , J. Math. Phys. **46** , 022104 (2005) .

[44] D.R.Herrick , J. Math. Phys. **16** , 281 (1975) ;

D.R.Herrick and F. H. Stillinger  Phys.Rev. **11** , 42 (1975) .

[45 ] M. Taut , K. Pernal, J. Cioslowski and V. Staemmler , J. Chem. Phys. **118** , 4861,(2003) .

[46 ] F. Holka, P. Neogrady ,V. Kello , M. Urban and G. H. F. Diercksen , Mol. Phys. **103**, 2740 (2005) . References therein.

**Figure Captions** :

**Fig. 1** :The schematic representation of the spherically confined model defined by the inner radius $R_a$ and the outer radius $R_b$ .The spherical boundaries correspond to the impenetrable walls located at the radial nodes of the chosen free (unconfined ) isotropic harmonic oscillator states given in Table 1 .

**Fig 2**:The plot of free isotropic harmonic oscillator radial density distribution function $4\pi r^2 [R_{4s}(r)]^2$ vs r . The three radial nodes are shown . The three confined states depicted on the left hand side define the confined isotropic harmonic oscillator states $1s_c$ , $2s_c$ , $3s_c$ which are obtained with common $R_a = 0$ but different $R_b (=R_c)$ , respectively , chosen as the first , second and third radial node . The other three confined states shown on the left side are obtained with $R_a (=R_c)$ , respectively , chosen at the first , second and third radial node and common $R_b = \infty$. All the six confined states have energy equal to the 4s state of the free isotropic harmonic oscillator which defines the incidental degeneracy condition.

**Fig. 3**: The variation of energy difference $\Delta E = E(n, \ell=2) - E(n+1, \ell=0)$ as a function of confinement radius $R_c$ corresponding to n=1 and n= 9 states of the isotropic harmonic oscillator. The curves cross at $R_c$ = 1.224745 a.u. , i.e. at the radial node of 2s state of the unconfined oscillator corresponding to which the $\Delta E = 2\hbar\omega$ as marked by an arrow.

**Fig. 4**:The spherically confined 10s and 9d isotropic harmonic oscillator states at $R_c$= 1.224745 ,i.e. the radial node of unconfined 2s state. The $\Delta E = 2\hbar\omega$ for the two confined states.

Table 1: A comparison of presently calculated energy of the confined isotropic harmonic oscillator with the previous most accurate calculations due to Aquino [29]. All values are in harmonic oscillator unit.

| State(n,ℓ) | $R_c$ (a.u) | Energy | |
|---|---|---|---|
| | | Present | Aquino [Ref. 29] |
| (0,0) 1s | 1  | 5.07558201560823  | 5.0755820152 |
|          | 2  | 1.76481643889110  | 1.7648164387 |
|          | 3  | 1.50608152728364  | 1.5060815272 |
|          | 4  | 1.50001460297369  | 1.5000146030 |
|          | 50 | 1.50000000000005  | 1.5000000000 |
| (0,1) 1p | 1  | 10.28225693949930 | 10.2822569390 |
|          | 2  | 3.24694709888081  | 3.2469470987 |
|          | 3  | 2.53129246671505  | 2.5312924666 |
|          | 4  | 2.50014377821833  | 2.5001437781 |
|          | 50 | 2.49999999999971  | 2.5000000000 |
| (0,2) 1d | 1  | 16.82777710958170 | 16.8277771098 |
|          | 2  | 5.01004086564171  | 5.0100408656 |
|          | 3  | 3.59824769903889  | 3.5982476989 |
|          | 4  | 3.50084207393092  | 3.5008420738 |
|          | 50 | 3.49999999999971  | 3.5000000000 |
| (1,0) 2s | 1  | 19.89969650042440 | 19.8996965018 |
|          | 2  | 5.58463907868235  | 5.5846390790 |
|          | 3  | 3.66421964490928  | 3.6642196450 |
|          | 4  | 3.50169153858314  | 3.5016915385 |
|          | 50 | 3.49999999999937  | 3.5000000000 |
| (1,1) 2p | 1  | 30.01348759147360 | 30.0134875924 |
|          | 2  | 8.15952888155243  | 8.1595288816 |
|          | 3  | 4.91389769057150  | 4.9138976907 |
|          | 4  | 4.50833043076121  | 4.5083304308 |
|          | 50 | 4.49999999999892  | 4.5000000000 |
| (2,0) 3s | 1  | 44.57717122947240 | 44.5771712285 |
|          | 2  | 11.76498212258770 | 11.7649821223 |
|          | 3  | 6.47333661637549  | 6.4733366162 |
|          | 4  | 5.53942179704881  | 5.5394217970 |
|          | 50 | 5.49999999999869  | 5.5000000000 |

Table 2: A comparison of energy values of confined H atom Calculated using the GPS method with the most accurate previous calculation of ref. [10].

| Level | $\ell$ | state | Rc | Energy Present | Ref. [10] |
|---|---|---|---|---|---|
| 1 | 0 | 1s | 1 | 2.373990866450 | 2.373990866 |
| 2 |  | 2s | 1 | 16.570256092140 | 16.570256093 |
| 5 |  | 5s | 1 | 119.327062496839 | 119.327062496 |
| 1 |  | 1s | 2 | -0.124999999938 | -0.125000000 |
| 2 |  | 2s | 2 | 3.327509156175 | 3.327509156 |
| 5 |  | 5s | 2 | 28.813505720908 | 28.813505721 |
| 1 |  | 1s | 20 | -0.500000000003 | -0.500000000 |
| 2 |  | 2s | 20 | -0.124987114308 | -0.124987114 |
| 5 |  | 5s | 20 | 0.112877739399 | 0.112877739 |
| 1 | 1 | 2p | 1 | 8.223138316542 | 8.223138316 |
| 4 |  | 5p | 1 | 95.991853334817 | 95.991853335 |
| 5 |  | 6p | 1 | 145.143445992318 | 145.143445993 |
| 1 |  | 2p | 2 | 1.576018785710 | 1.576018786 |
| 4 |  | 5p | 2 | 23.259082803917 | 23.259082804 |
| 5 |  | 6p | 2 | 35.498705803768 | 35.498705804 |
| 1 |  | 2p | 20 | -0.124994606647 | -0.124994607 |
| 4 |  | 5p | 20 | 0.095169727184 | 0.095169727 |
| 5 |  | 6p | 20 | 0.209413062653 | 0.209413063 |

Table 3: The numerically estimates of radial nodes (in a.u.) calculated from the spherical Hermite polynomials describing the wave functions of the free ($R_c = \infty$) isotropic harmonic oscillator up to n = 3 and $\ell$ =3. The first node for (1, $\ell$) state is given by $\{(2\ell+3)/2\}^{1/2}$ a.u. when $V(r) = r^2/2$.

| (n,$\ell$) | $R_{node}(1)$ | $R_{node}(2)$ | $R_{node}(3)$ |
|---|---|---|---|
| (1,0) [2s] | 1.224745 |  |  |
| (1,1) [2p] | 1.581139 |  |  |
| (1,2) [2d] | 1.870829 |  |  |
| (1,3) [2f] | 2.121320 |  |  |
| (2,0) [3s] | 0.958572 | 2.020183 |  |
| (2,1) [3p] | 1.276390 | 2.317505 |  |
| (2,2) [3d] | 1.542296 | 2.573192 |  |
| (2,3) [3f] | 1.776173 | 2.800929 |  |
| (3,0) [4s] | 0.816288 | 1.673552 | 2.651961 |
| (3,1) [4p] | 1.104718 | 1.951635 | 2.910450 |
| (3,2) [4d] | 1.350859 | 2.194025 | 3.140292 |
| (3,3) [4f] | 1.570165 | 2.411845 | 3.349267 |

Table 4: Incidental degeneracy (*) of the $\ell=0$ states of the spherically confined isotropic harmonic oscillator in the first four levels (n=1-4). The energy under confinement equals that of the unconfined state shown under column 5. The initial and final values of the confining boundaries are given as $R_a$ and $R_b$ (Fig. 1) in a. u. Energy values are given in harmonic oscillator units with $V(r) = r^2/2$ a.u.

| Confined (n,ℓ) | $R_a$ | $R_b$ | Unconfined (n,ℓ) | E |
|---|---|---|---|---|
| (0,0) | 0 | 200 | | 1.5 |
| (1,0) | 0 | 200 | | 3.5 |
| (2,0) | 0 | 200 | | 5.5 |
| (3,0) | 0 | 200 | | 7.5 |
| | | | | |
| (0,0) | 0 | 1.22474 | (1,0) | 3.50000* |
| (1,0) | 0 | 1.22474 | | 13.40049 |
| (2,0) | 0 | 1.22474 | | 29.85493 |
| (3,0) | 0 | 1.22474 | | 52.88573 |
| | | | | |
| (0,0) | 1.22474 | 200 | (1,0) | 3.50000* |
| (1,0) | 1.22474 | 200 | | 6.192747 |
| (2,0) | 1.22474 | 200 | | 8.70989 |
| (3,0) | 1.22474 | 200 | | 11.14072 |
| | | | | |
| (0,0) | 0 | 0.95857 | (2,0) | 5.50000* |
| (1,0) | 0 | 0.95857 | | 21.62972 |
| (2,0) | 0 | 0.95857 | | 48.48575 |
| (3,0) | 0 | 0.95857 | | 86.08084 |
| | | | | |
| (0,0) | 0 | 2.02018 | | 1.7497 |
| (1,0) | 0 | 2.02018 | (2,0) | 5.50000* |
| (2,0) | 0 | 2.02018 | | 11.55791 |
| (3,0) | 0 | 2.02018 | | 20.02472 |
| | | | | |
| (0,0) | 0.95857 | 200 | | 2.95671 |
| (1,0) | 0.95857 | 200 | (2,0) | 5.50000* |
| (2,0) | 0.95857 | 200 | | 7.90539 |
| (3,0) | 0.95857 | 200 | | 10.24301 |
| | | | | |
| (0,0) | 2.02018 | 200 | (2,0) | 5.50000* |
| (1,0) | 2.02018 | 200 | | 8.63097 |
| (2,0) | 2.02018 | 200 | | 11.47792 |
| (3,0) | 2.02018 | 200 | | 14.18472 |
| | | | | |
| (0,0) | 0 | 0.81629 | (3,0) | 7.50000* |

| Confined (n,ℓ) | $R_a$ | $R_b$ | Unconfined (n,ℓ) | E |
|---|---|---|---|---|
| (1,0) | 0 | 0.81629 | | 29.73084 |
| (2,0) | 0 | 0.81629 | | 66.76311 |
| (3,0) | 0 | 0.81629 | | 118.6058 |
| | | | | |
| (0,0) | 0 | 1.67355 | | 2.145774 |
| (1,0) | 0 | 1.67355 | (3,0) | 7.50000* |
| (2,0) | 0 | 1.67355 | | 16.31855 |
| (3,0) | 0 | 1.67355 | | 28.65478 |
| | | | | |
| (0,0) | 0 | 2.65196 | | 1.52851 |
| (1,0) | 0 | 2.65196 | | 3.96289 |
| (2,0) | 0 | 2.65196 | (3,0) | 7.50000* |
| (3,0) | 0 | 2.65196 | | 12.4094 |
| | | | | |
| (0,0) | 0.81629 | 200 | | 2.69161 |
| (1,0) | 0.81629 | 200 | | 5.15456 |
| (2,0) | 0.81629 | 200 | (3,0) | 7.50000* |
| (3,0) | 0.81629 | 200 | | 9.78765 |
| | | | | |
| (0,0) | 1.67355 | 200 | | 4.558302 |
| (1,0) | 1.67355 | 200 | (3,0) | 7.50000* |
| (2,0) | 1.67355 | 200 | | 10.20407 |
| (3,0) | 1.67355 | 200 | | 12.79114 |
| | | | | |
| (0,0) | 2.65196 | 200 | (3,0) | 7.50000* |
| (1,0) | 2.65196 | 200 | | 10.96875 |
| (2,0) | 2.65196 | 200 | | 14.07236 |
| (3,0) | 2.65196 | 200 | | 16.99494 |

Table 5. Incidental degeneracy (*) of the $\ell=3$ states of the spherically confined isotropic harmonic oscillator in the first four levels (n=1-4). The energy under confinement equals that of the unconfined state shown under column 5. The initial and final values of the confining boundaries are given as $R_a$ and $R_b$ (Fig. 1) in a. u. Energy values are given in harmonic oscillator units with $V(r) = r^2/2$ a.u.

| Confined (n,ℓ) | $R_a$ | $R_b$ | Unconfined (n,ℓ) | E |
|---|---|---|---|---|
| (0,3) | 0 | 200 | | 4.5 |
| (1,3) | 0 | 200 | | 6.5 |
| (2,3) | 0 | 200 | | 8.5 |
| (3,3) | 0 | 200 | | 10.5 |
| | | | | |
| (0,3) | 0 | 2.12132 | (1,3) | 6.5* |
| (1,3) | 0 | 2.12132 | | 12.96371 |
| (2,3) | 0 | 2.12132 | | 21.69090 |
| (3,3) | 0 | 2.12132 | | 32.63444 |

| | | | | | | | |
|---|---|---|---|---|---|---|---|
| (0,3) | 2.12132 | | 200 | (1,3) | 6.5* | | |
| (1,3) | 2.12132 | | 200 | | 9.52558 | | |
| (2,3) | 2.12132 | | 200 | | 12.33182 | | |
| (3,3) | 2.12132 | | 200 | | 15.02062 | | |
| | | | | | | | |
| (0,3) | 0 | | 1.77617 | (2,3) | 8.50000* | | |
| (1,3) | 0 | | 1.77617 | | 17.83374 | | |
| (2,3) | 0 | | 1.77617 | | 30.32795 | | |
| (3,3) | 0 | | 1.77617 | | 45.96060 | | |
| | | | | | | | |
| (0,3) | 0 | | 2.80093 | | 4.90736 | | |
| (1,3) | 0 | | 2.80093 | (2,3) | 8.50000* | | |
| (2,3) | 0 | | 2.80093 | | 13.43698 | | |
| (3,3) | 0 | | 2.80093 | | 19.67649 | | |
| | | | | | | | |
| (0,3) | 1.77617 | | 200 | | 5.71300 | | |
| (1,3) | 1.77617 | | 200 | (2,3) | 8.50000* | | |
| (2,3) | 1.77617 | | 200 | | 11.14011 | | |
| (3,3) | 1.77617 | | 200 | | 13.69521 | | |
| | | | | | | | |
| (0,3) | 2.80093 | | 200 | (2,3) | 8.50000* | | |
| (1,3) | 2.80093 | | 200 | | 11.95128 | | |
| (2,3) | 2.80093 | | 200 | | 15.06296 | | |

Table 6: Constant splitting $\Delta E = 2\hbar\omega$, of the pairs of isotropic harmonic oscillator confined at the radius $R_a = 0$ and $R_b = \{(2\ell+3)/2\}^{1/2}$ a.u. The first set of pairs under columns 5 and 6 correspond to the [(2s,1d), (3s,2d) ..] or [{(1,0), (0,2)}, {(2,0), (1,2)} ..] pairs and those given in the last two columns correspond to the [(2h,1j), (3h,2j) ..] or [{(1,5), (0,7)}, {(2,5), (1,7)} ..] pairs. In each case the energy difference is exactly two oscillator units. In the free state all pairs correspond to degenerate energy.

| n | ℓ | $R_a$ | $R_b$ | E(n+1,ℓ) | E(n,ℓ+2) | ℓ | $R_b$ | E(n+1,ℓ) | E(n,ℓ+2) |
|---|---|---|---|---|---|---|---|---|---|
| 0 | 0 | 0 | 1.22474 | 11.40049 | 13.40049 | 5 | 2.54951 | 12.38991 | 14.38991 |
| 1 | 0 | 0 | 1.22474 | 27.85493 | 29.85493 | 5 | 2.54951 | 19.89793 | 21.89793 |
| 2 | 0 | 0 | 1.22474 | 50.88573 | 52.88573 | 5 | 2.54951 | 28.96223 | 30.96223 |
| 3 | 0 | 0 | 1.22474 | 80.49532 | 82.49532 | 5 | 2.54951 | 39.56283 | 41.56283 |
| 4 | 0 | 0 | 1.22474 | 116.6843 | 118.6843 | 5 | 2.54951 | 51.69147 | 53.69147 |
| 5 | 0 | 0 | 1.22474 | 159.4528 | 161.4528 | 5 | 2.54951 | 65.34419 | 67.34419 |
| 6 | 0 | 0 | 1.22474 | 208.8010 | 210.8010 | 5 | 2.54951 | 80.51888 | 82.51888 |
| 7 | 0 | 0 | 1.22474 | 264.7289 | 266.7289 | 5 | 2.54951 | 97.21433 | 99.21433 |
| 8 | 0 | 0 | 1.22474 | 327.2365 | 329.2365 | 5 | 2.54951 | 115.4298 | 117.4298 |
| 9 | 0 | 0 | 1.22474 | 396.3238 | 398.3238 | 5 | 2.54951 | 135.1648 | 137.1648 |
| 39 | 0 | 0 | 1.22474 | 5528.518 | 5530.518 | 5 | 2.54951 | 1433.375 | 1435.375 |
| 79 | 0 | 0 | 1.22474 | 21583.07 | 21585.07 | 5 | 2.54951 | 5290.111 | 5292.111 |
| 99 | 0 | 0 | 1.22474 | 33558.19 | 33560.19 | 5 | 2.54951 | 8129.520 | 8131.520 |

Table 7 : Incidental degeneracy of the Davidson oscillator and the ordering of [(1,0), (0,2) ] like first nine pairs under spherical confinement at $R_c$ , corresponding to the node of free (1,0) state. At this radius , the confined (0,0) level corresponds to the energy of the free (1,0) level shown by asterisk (*). The last column gives the difference of energy of the successive pairs which is found to decrease slowly with increasing principal quantum number ,n .

| (n,ℓ) | $E^{Free}$ [ℓ=0] | $R_c$ | $E(R_c)$[ℓ=0] | (n,ℓ') | $E(R_c)$[ℓ'=2] | ΔE( ℓ-ℓ' ) | ΔΔE |
|---|---|---|---|---|---|---|---|
|  | 2.118034 | 1.45535 | 4.118034* |  |  |  |  |
| (1,0) | 4.118034 | 1.45535 | 12.54406 | (0,2) | 8.9773544 | 3.566702 |  |
| (2,0) | 6.118034 | 1.45535 | 25.63112 | (1,2) | 21.053799 | 4.577321 | 1.010619 |
| (3,0) | 8.118034 | 1.45535 | 43.37929 | (2,2) | 37.804514 | 5.574771 | 0.997450 |
| (4,0) | 10.118034 | 1.45535 | 65.78776 | (3,2) | 59.218965 | 6.568799 | 0.994028 |
| (5,0) | 12.118034 | 1.45535 | 92.85627 | (4,2) | 85.294735 | 7.561537 | 0.992738 |
| (6,0) | 14.118034 | 1.45535 | 124.5847 | (5,2) | 116.03101 | 8.553680 | 0.992143 |
| (7,0) | 16.118034 | 1.45535 | 160.9730 | (6,2) | 151.42747 | 9.545509 | 0.991829 |
| (8,0) | 18.118034 | 1.45535 | 202.0211 | (7,2) | 191.48394 | 10.53716 | 0.991649 |
| (9,0) | 20.118034 | 1.45535 | 247.7290 | (8,2) | 236.20035 | 11.52869 | 0.991537 |

Table 8: Spherically confined harmonic oscillator with penetrable potential barrier $V(R_c)$ , located at $R_c$ . E' , E" , and E'" correspond to the locations 1.22511 , 1.99975 and 4.00052 a.u. , respectively. The first five ℓ=0 states are shown corresponding to each potential barrier. All values are given in a.u.

| n | $V(R_c)$ | E' | E" | E'" |
|---|---|---|---|---|
| 0 | 100 | 3.231296 | 1.72558 | 1.50001 |
| 1 | 100 | 12.24971 | 5.36078 | 3.50126 |
| 2 | 100 | 27.15094 | 11.20873 | 5.53153 |
| 3 | 100 | 47.79611 | 19.36623 | 7.75104 |
| 4 | 100 | 73.75508 | 29.82322 | 10.42447 |
| 0 | 50 | 3.10622 | 1.70612 | 1.50001 |
| 1 | 50 | 11.68400 | 5.24428 | 3.50103 |
| 2 | 50 | 25.67700 | 10.90704 | 5.52694 |
| 3 | 50 | 44.19430 | 18.77119 | 7.72434 |
| 4 | 50 | 50.01406 | 28.76725 | 10.35244 |
| 0 | 50 | 2.63346 | 1.63055 | 1.50001 |
| 1 | 10 | 8.97151 | 4.72514 | 3.50014 |
| 2 | 10 | 10.01446 | 9.18980 | 5.50443 |
| 3 | 10 | 10.05777 | 10.00052 | 7.54548 |
| 4 | 10 | 10.12977 | 10.00207 | 9.6295 |
| 0 | 5 | 2.31685 | 1.57677 | 1.499997 |
| 1 | 5 | 5.01364 | 4.20321 | 3.49936 |
| 2 | 5 | 5.05452 | 5.00052 | 5.00053 |
| 3 | 5 | 5.12253 | 5.00207 | 5.00212 |
| 4 | 5 | 5.21749 | 5.00464 | 5.00476 |

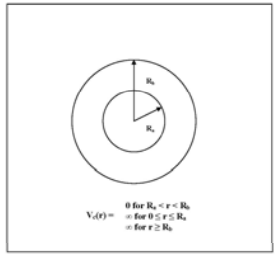

Fig.1

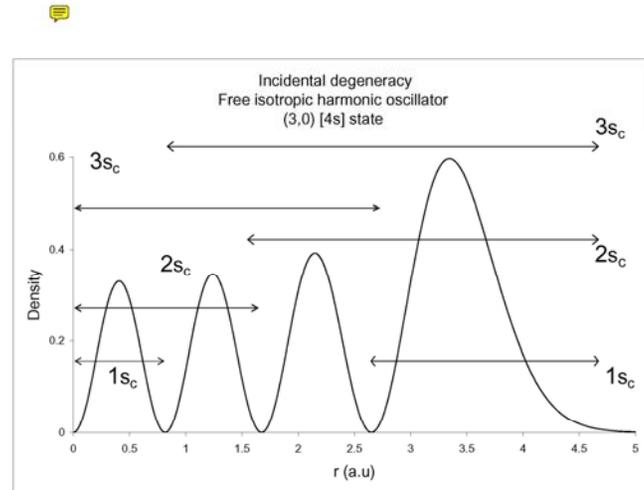

Fig. 2

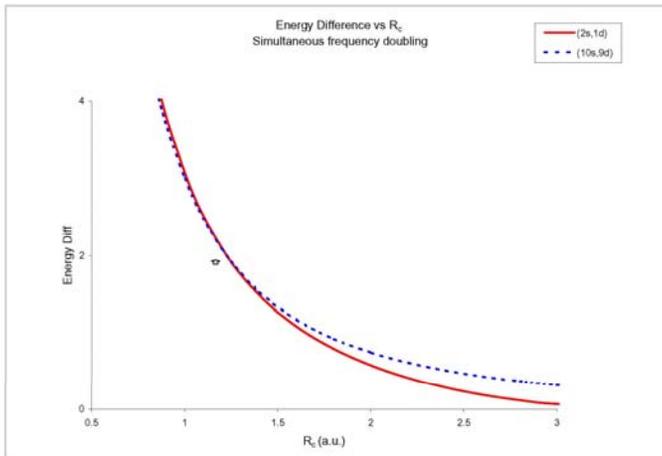

Fig. 3

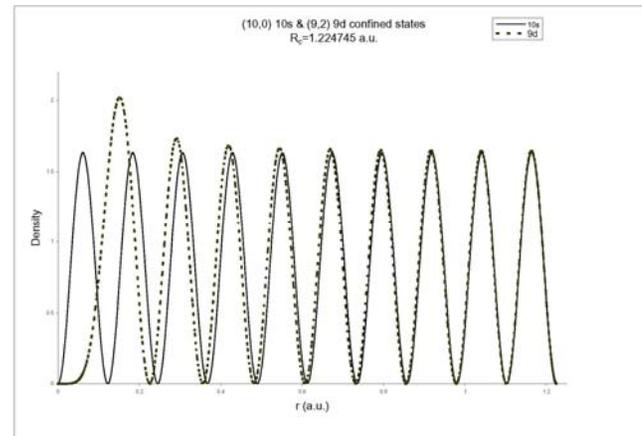

Fig. 4